\def\journal{\topmargin .3in	\oddsidemargin .5in
	\headheight 0pt	\headsep 0pt
	\textwidth 5.625in 
	\textheight 8.25in 
	\marginparwidth 1.5in
	\parindent 2em
	\parskip .5ex plus .1ex		\jot = 1.5ex}
\documentstyle[12pt]{article}
\journal

\catcode`\@=11
\def\marginnote#1{}
\newcount\hour
\newcount\minute
\newtoks\amorpm
\hour=\time\divide\hour by60
\minute=\time{\multiply\hour by60 \global\advance\minute by-\hour}
\edef\standardtime{{\ifnum\hour<12 \global\amorpm={am}%
	\else\global\amorpm={pm}\advance\hour by-12 \fi
	\ifnum\hour=0 \hour=12 \fi
	\number\hour:\ifnum\minute<10 0\fi\number\minute\the\amorpm}}
\edef\militarytime{\number\hour:\ifnum\minute<10 0\fi\number\minute}
\def\draftlabel#1{{\@bsphack\if@filesw {\let\thepage\relax
   \xdef\@gtempa{\write\@auxout{\string
      \newlabel{#1}{{\@currentlabel}{\thepage}}}}}\@gtempa
   \if@nobreak \ifvmode\nobreak\fi\fi\fi\@esphack}
	\gdef\@eqnlabel{#1}}
\def\@eqnlabel{}
\def\@vacuum{}
\def\draftmarginnote#1{\marginpar{\raggedright\scriptsize\tt#1}}
\def\draft{\oddsidemargin -.5truein
	\def\@oddfoot{\sl preliminary draft \hfil
	\rm\thepage\hfil\sl\today\quad\militarytime}
	\let\@evenfoot\@oddfoot	\overfullrule 3pt
	\let\label=\draftlabel
	\let\marginnote=\draftmarginnote
   \def\@eqnnum{(\theequation)\rlap{\kern\marginparsep\tt\@eqnlabel}%
\global\let\@eqnlabel\@vacuum}  }
\def\preprint{\twocolumn\sloppy\flushbottom\parindent 2em
	\leftmargini 2em\leftmarginv .5em\leftmarginvi .5em
	\oddsidemargin -.5in	\evensidemargin -.5in
	\columnsep .4in	\footheight 0pt
	\textwidth 10in	\topmargin  -.4in
	\headheight 12pt \topskip .4in
	\textheight 7.1in \footskip 0pt
	\def\@oddhead{\thepage\hfil\addtocounter{page}{1}\thepage}
	\let\@evenhead\@oddhead	\def\@oddfoot{}	\def\@evenfoot{} }
\def\numberbysection{\@addtoreset{equation}{section}
	\def\theequation{\thesection.\arabic{equation}}}
\def\underline#1{\relax\ifmmode\@@underline#1\else
	$\@@underline{\hbox{#1}}$\relax\fi}
\def\titlepage{\@restonecolfalse\if@twocolumn\@restonecoltrue\onecolumn
     \else \newpage \fi \thispagestyle{empty}\c@page\z@
	\def\thefootnote{\fnsymbol{footnote}} }
\def\endtitlepage{\if@restonecol\twocolumn \else \newpage \fi
	\def\thefootnote{\arabic{footnote}}
	\setcounter{footnote}{0}}  
\catcode`@=12
\relax
\def\figcap{\section*{Figure Captions\markboth
	{FIGURECAPTIONS}{FIGURECAPTIONS}}\list
	{Figure \arabic{enumi}:\hfill}{\settowidth\labelwidth{Figure 999:}
	\leftmargin\labelwidth
	\advance\leftmargin\labelsep\usecounter{enumi}}}
 \relax
\def\tablecap{\section*{Table Captions\markboth
	{TABLECAPTIONS}{TABLECAPTIONS}}\list
	{Table \arabic{enumi}:\hfill}{\settowidth\labelwidth{Table 999:}
	\leftmargin\labelwidth
	\advance\leftmargin\labelsep\usecounter{enumi}}}
 \relax
\def\reflist{\section*{References\markboth
	{REFLIST}{REFLIST}}\list
	{[\arabic{enumi}]\hfill}{\settowidth\labelwidth{[999]}
	\leftmargin\labelwidth
	\advance\leftmargin\labelsep\usecounter{enumi}}}
 \relax
\makeatletter
\newcounter{pubctr}
\def\publist{\@ifnextchar[{\@publist}{\@@publist}}
\def\@publist[#1]{\list
	{[\arabic{pubctr}]\hfill}{\settowidth\labelwidth{[999]}
	\leftmargin\labelwidth
	\advance\leftmargin\labelsep
	\@nmbrlisttrue\def\@listctr{pubctr}
	\setcounter{pubctr}{#1}\addtocounter{pubctr}{-1}}}
\def\@@publist{\list
	{[\arabic{pubctr}]\hfill}{\settowidth\labelwidth{[999]}
	\leftmargin\labelwidth
	\advance\leftmargin\labelsep
	\@nmbrlisttrue\def\@listctr{pubctr}}}
 \relax
\makeatother
\catcode`\@=11
\def\section{\@startsection {section}{1}{0pt}{-3.5ex plus -1ex minus
 -.2ex}{2.3ex plus .2ex}{\raggedright\large\bf}}
\catcode`\@=12
\newskip\humongous \humongous=0pt plus 1000pt minus 1000pt
\def\caja{\mathsurround=0pt}

\newif\ifdtup
\def\panorama{\global\dtuptrue \openup1\jot \caja
	\everycr{\noalign{\ifdtup \global\dtupfalse
	\vskip-\lineskiplimit \vskip\normallineskiplimit
	\else \penalty\interdisplaylinepenalty \fi}}}
\def\eqalignno#1{\panorama \tabskip=\humongous
	\halign to\displaywidth{\hfil$\displaystyle{##}$
	\tabskip=0pt&$\displaystyle{{}##}$\hfil
	\tabskip=\humongous&\llap{$##$}\tabskip=0pt
	\crcr#1\crcr}}

\def\oldreffmt#1{\rlap{[#1]} \hbox to 2\parindent{}}

\def\figfmt#1{\rlap{Figure {#1}} \hbox to 1in{}}

\def\beq{\begin{equation}}
\def\eeq{\end{equation}}

\def\bea{\begin{eqnarray}}

\def\eea{\end{eqnarray}}
\catcode`\@=11
\def\eqnarray{\stepcounter{equation}\let\@currentlabel=\theequation
\global\@eqnswtrue
\global\@eqcnt\z@\tabskip\@centering\let\\=\@eqncr
\gdef\@@fix{}\def\eqno##1{\gdef\@@fix{##1}}%
$$\halign to \displaywidth\bgroup\@eqnsel\hskip\@centering
  $\displaystyle\tabskip\z@{##}$&\global\@eqcnt\@ne
  \hskip 2\arraycolsep \hfil${##}$\hfil
  &\global\@eqcnt\tw@ \hskip 2\arraycolsep $\displaystyle\tabskip\z@{##}$\hfil
   \tabskip\@centering&\llap{##}\tabskip\z@\cr}

\def\@@eqncr{\let\@tempa\relax
    \ifcase\@eqcnt \def\@tempa{& & &}\or \def\@tempa{& &}
      \else \def\@tempa{&}\fi
     \@tempa \if@eqnsw\@eqnnum\stepcounter{equation}\else\@@fix\gdef\@@fix{}\fi
     \global\@eqnswtrue\global\@eqcnt\z@\cr}

\catcode`\@=12
\font\tenbifull=cmmib10 
\font\tenbimed=cmmib10 scaled 800
\font\tenbismall=cmmib10 scaled 666
\relax

\journal

\catcode`\@=11
\def\marginnote#1{}
\newcount\hour
\newcount\minute
\newtoks\amorpm
\hour=\time\divide\hour by60
\minute=\time{\multiply\hour by60 \global\advance\minute by-\hour}
\edef\standardtime{{\ifnum\hour<12 \global\amorpm={am}%
	\else\global\amorpm={pm}\advance\hour by-12 \fi
	\ifnum\hour=0 \hour=12 \fi
	\number\hour:\ifnum\minute<10 0\fi\number\minute\the\amorpm}}
\edef\militarytime{\number\hour:\ifnum\minute<10 0\fi\number\minute}
\def\draftlabel#1{{\@bsphack\if@filesw {\let\thepage\relax
   \xdef\@gtempa{\write\@auxout{\string
      \newlabel{#1}{{\@currentlabel}{\thepage}}}}}\@gtempa
   \if@nobreak \ifvmode\nobreak\fi\fi\fi\@esphack}
	\gdef\@eqnlabel{#1}}
\def\@eqnlabel{}
\def\@vacuum{}
\def\draftmarginnote#1{\marginpar{\raggedright\scriptsize\tt#1}}
\def\draft{\oddsidemargin -.5truein
	\def\@oddfoot{\sl preliminary draft \hfil
	\rm\thepage\hfil\sl\today\quad\militarytime}
	\let\@evenfoot\@oddfoot	\overfullrule 3pt
	\let\label=\draftlabel
	\let\marginnote=\draftmarginnote
   \def\@eqnnum{(\theequation)\rlap{\kern\marginparsep\tt\@eqnlabel}%
\global\let\@eqnlabel\@vacuum}  }
\def\preprint{\twocolumn\sloppy\flushbottom\parindent 2em
	\leftmargini 2em\leftmarginv .5em\leftmarginvi .5em
	\oddsidemargin -.5in	\evensidemargin -.5in
	\columnsep .4in	\footheight 0pt
	\textwidth 10in	\topmargin  -.4in
	\headheight 12pt \topskip .4in
	\textheight 7.1in \footskip 0pt
	\def\@oddhead{\thepage\hfil\addtocounter{page}{1}\thepage}
	\let\@evenhead\@oddhead	\def\@oddfoot{}	\def\@evenfoot{} }
\def\numberbysection{\@addtoreset{equation}{section}
	\def\theequation{\thesection.\arabic{equation}}}
\def\underline#1{\relax\ifmmode\@@underline#1\else
	$\@@underline{\hbox{#1}}$\relax\fi}
\def\titlepage{\@restonecolfalse\if@twocolumn\@restonecoltrue\onecolumn
     \else \newpage \fi \thispagestyle{empty}\c@page\z@
	\def\thefootnote{\fnsymbol{footnote}} }
\def\endtitlepage{\if@restonecol\twocolumn \else \newpage \fi
	\def\thefootnote{\arabic{footnote}}
	\setcounter{footnote}{0}}  
\catcode`@=12
\relax
\def\figcap{\section*{Figure Captions\markboth
	{FIGURECAPTIONS}{FIGURECAPTIONS}}\list
	{Figure \arabic{enumi}:\hfill}{\settowidth\labelwidth{Figure 999:}
	\leftmargin\labelwidth
	\advance\leftmargin\labelsep\usecounter{enumi}}}
 \relax
\def\tablecap{\section*{Table Captions\markboth
	{TABLECAPTIONS}{TABLECAPTIONS}}\list
	{Table \arabic{enumi}:\hfill}{\settowidth\labelwidth{Table 999:}
	\leftmargin\labelwidth
	\advance\leftmargin\labelsep\usecounter{enumi}}}
 \relax
\def\reflist{\section*{References\markboth
	{REFLIST}{REFLIST}}\list
	{[\arabic{enumi}]\hfill}{\settowidth\labelwidth{[999]}
	\leftmargin\labelwidth
	\advance\leftmargin\labelsep\usecounter{enumi}}}
 \relax
\makeatletter
\newcounter{pubctr}
\def\publist{\@ifnextchar[{\@publist}{\@@publist}}
\def\@publist[#1]{\list
	{[\arabic{pubctr}]\hfill}{\settowidth\labelwidth{[999]}
	\leftmargin\labelwidth
	\advance\leftmargin\labelsep
	\@nmbrlisttrue\def\@listctr{pubctr}
	\setcounter{pubctr}{#1}\addtocounter{pubctr}{-1}}}
\def\@@publist{\list
	{[\arabic{pubctr}]\hfill}{\settowidth\labelwidth{[999]}
	\leftmargin\labelwidth
	\advance\leftmargin\labelsep
	\@nmbrlisttrue\def\@listctr{pubctr}}}
 \relax
\makeatother
\catcode`\@=11
\def\section{\@startsection {section}{1}{0pt}{-3.5ex plus -1ex minus
 -.2ex}{2.3ex plus .2ex}{\raggedright\large\bf}}
\catcode`\@=12
\newskip\humongous \humongous=0pt plus 1000pt minus 1000pt
\def\caja{\mathsurround=0pt}

\newif\ifdtup
\def\panorama{\global\dtuptrue \openup1\jot \caja
	\everycr{\noalign{\ifdtup \global\dtupfalse
	\vskip-\lineskiplimit \vskip\normallineskiplimit
	\else \penalty\interdisplaylinepenalty \fi}}}
\def\eqalignno#1{\panorama \tabskip=\humongous
	\halign to\displaywidth{\hfil$\displaystyle{##}$
	\tabskip=0pt&$\displaystyle{{}##}$\hfil
	\tabskip=\humongous&\llap{$##$}\tabskip=0pt
	\crcr#1\crcr}}

\def\oldreffmt#1{\rlap{[#1]} \hbox to 2\parindent{}}

\def\figfmt#1{\rlap{Figure {#1}} \hbox to 1in{}}

\def\beq{\begin{equation}}
\def\eeq{\end{equation}}

\def\bea{\begin{eqnarray}}

\def\eea{\end{eqnarray}}
\catcode`\@=11
\def\eqnarray{\stepcounter{equation}\let\@currentlabel=\theequation
\global\@eqnswtrue
\global\@eqcnt\z@\tabskip\@centering\let\\=\@eqncr
\gdef\@@fix{}\def\eqno##1{\gdef\@@fix{##1}}%
$$\halign to \displaywidth\bgroup\@eqnsel\hskip\@centering
  $\displaystyle\tabskip\z@{##}$&\global\@eqcnt\@ne
  \hskip 2\arraycolsep \hfil${##}$\hfil
  &\global\@eqcnt\tw@ \hskip 2\arraycolsep $\displaystyle\tabskip\z@{##}$\hfil
   \tabskip\@centering&\llap{##}\tabskip\z@\cr}

\def\@@eqncr{\let\@tempa\relax
    \ifcase\@eqcnt \def\@tempa{& & &}\or \def\@tempa{& &}
      \else \def\@tempa{&}\fi
     \@tempa \if@eqnsw\@eqnnum\stepcounter{equation}\else\@@fix\gdef\@@fix{}\fi
     \global\@eqnswtrue\global\@eqcnt\z@\cr}

\catcode`\@=12
\font\tenbifull=cmmib10 
\font\tenbimed=cmmib10 scaled 800
\font\tenbismall=cmmib10 scaled 666
\relax

\def\thefootnote{\fnsymbol{footnote}}
\def\ref#1{$^{#1)}$}
\begin{document}
\begin{titlepage}
\begin{center}

\today     \hfill    LBL-32484 \\
          \hfill    UCB-PTH-92/21 \\
          \hfill   OHSTPY-HEP-T-92-015\\
          \hfill   Stanford ITP 922/92\\
\vskip .20in

{\large \bf Neutrino Masses and Mixing Angles in a

Predictive Theory of Fermion Masses }
\footnote{This work was supported in part by the Director, Office of
Energy Research, Office of High Energy and Nuclear Physics, Division of
High Energy Physics of the U.S. Department of Energy under Contracts
DE-AC03-76SF00098 and DOE-ER-01545-585 and in part by the National Science
Foundation under grants PHY90-21139 and PHY86-12280.}
\vskip .20in

Savas Dimopoulos\\[.20in]

{\em Department of Physics,
Stanford University,
Stanford, CA 94305}\\[.20in]

Lawrence J. Hall\\[.20in]

{\em  Department of Physics and Lawrence Berkeley Laboratory,\\
  University of California, Berkeley, California 94720}\\[.20in]
Stuart Raby\\[.20in]
{\em Department of Physics,
The Ohio State University,
Columbus, OH 43210}
\end{center}

\vskip .10in

\begin{abstract}
A framework for predicting charged fermion masses in supersymmetric grand
unified theories is extended to make predictions in the neutrino sector.
Eight new predictions are made: the two neutrino mass ratios and the three
mixing angles and three phases of the weak leptonic mixing matrix.
There are three versions of the theory which are relevant for
producing MSW neutrino oscillations in the sun.  One of these is
prefered by the combined solar neutrino observations. Another will be probed
significantly by the searches for $\nu_\mu\nu_\tau$ oscillations at the
NOMAD, CHORUS and P803 experiments. In this second
version $\nu_\tau$ could be a significant component of
the dark matter in the universe.
\end{abstract}
\end{titlepage}
PAGE (PAGE 1) \renewcommand{\thepage}{\roman{page}} \setcounter{page}{2}
\mbox{ }

\vskip 1in

\begin{center}
{\bf Disclaimer}
\end{center}

\vskip .2in

\begin{scriptsize}
\begin{quotation}
This document was prepared as an account of work sponsored by the United
States Government.  Neither the United States Government nor any agency
thereof, nor The Regents of the University of California, nor any of their
employees, makes any warranty, express or implied, or assumes any legal
liability or responsibility for the accuracy, completeness, or usefulness
of any information, apparatus, product, or process disclosed, or represents
that its use would not infringe privately owned rights.  Reference herein
to any specific commercial products process, or service by its trade name,
trademark, manufacturer, or otherwise, does not necessarily constitute or
imply its endorsement, recommendation, or favoring by the United States
Government or any agency thereof, or The Regents of the University of
California.  The views and opinions of authors expressed herein do not
necessarily state or reflect those of the United States Government or any
agency thereof of The Regents of the University of California and shall
not be used for advertising or product endorsement purposes.
\end{quotation}
\end{scriptsize}

\vskip 2in

\begin{center}
\begin{small}
{\it Lawrence Berkeley Laboratory is an equal opportunity employer.}
\end{small}
\end{center}

\newpage
\renewcommand{\thepage}{\arabic{page}}
\setcounter{page}{1}

 %


The solar neutrino problem and the several existing and upcoming
neutrino experiments have caused interest in neutrino masses and
mixings to escalate in the last few years.
Theoretical advances on this subject are very difficult to come by since the
subject of neutrino masses is, in general, coupled to the problem
of quark and charged lepton masses on which very little progress
has been made.

Recently, we proposed a predictive framework, based on
supersymmetric grand unified theories (GUTs), in which the
14 parameters of the quark and charged lepton mass matrices,
plus the ratio of the Higgs vacuum expectation values (vevs), can
be obtained in terms of just 8 input parameters, thus
leading to 6 predictions \cite{dhr}.
The consequences of these predictions will be tested in planned experiments
\cite{barger}.
Our framework has
the virtue that it is a consequence of a large class of GUTs.
In this paper we wish to study a subset of models
within this class which are very predictive in the neutrino
sector. We will show how the addition of one more input parameter,
for a total of 9 inputs, allows us to account for 23 parameters,
resulting in 14 predictions.  In particular we will predict the 2
neutrino mass ratios, 3 mixing angles and 3 phases of the lepton
sector.

We begin with a rapid overview of our framework.  A key ingredient
is the Georgi-Jarlskog texture for fermion mass matrices at the
GUT scale \cite{georgi}:

$$M_D = \left( \begin{array}{ccc}
0 & F & 0\\
F & E & 0\\
0 & 0 & D
\end{array} \right){v \over \sqrt{2}}\cos\beta
\;\;\; \; \; \; \;
M_E = \left( \begin{array}{ccc}
0 & F & 0\\
F & -3E &0 \\
0 & 0 & D
\end{array} \right){v \over \sqrt{2}}\cos\beta $$
\\
$$M_U = \left( \begin{array}{ccc}
0 & C & 0\\
C & 0 & B\\
0 & B & A
\end{array} \right){v \over \sqrt{2}}\sin\beta \eqno (1)
$$
where tan $\beta = v_2/v_1$, is the ratio of electroweak breaking
vevs. The factor of 3 difference in the 22 elements of $M_D$ and
$M_E$ is of crucial importance.  It arises naturally as a
consequence of the breaking of the Pati-Salam SU(4) via a vev
pointing parallel to the hypercharge generator

$$Y_{15} = \left( \begin{array}{cccc}
1 & & & \\
 & 1 & & \\
 & & 1 & \\
&&&-3
\end{array} \right) \eqno (2)
$$
Three examples where this happens
are when the 22 entries are generated by Higgs doublets which lie
in

(a)  a $\overline{45}$ of SU(5),

(b)  a $\overline{126}$ of SO(10),

(c) a higher dimension operator, for example 45 $\times$ 45 $\times$ 10 of
SO(10).

For simplicity and definiteness we will focus on case (b) in this
paper and thus study theories where the $M_D$ and $M_E$ arise from
the following matrix of Yukawa
couplings.\footnote{Many of our conclusions are
probably more general and valid even if fermion masses come from
higher dimension operators.}

$$
16 \left( \begin{array}{ccc}
0 & f\, 10^d & 0\\[5pt]
 f\, 10^d  & e\, \overline{126}^d   & 0 \\[5pt]
0 & 0 & d\,10^d
\end{array} \right)16 \eqno (3)
$$
Here $d$, $e$ and $f$ are Yukawa couplings and $10^d$ and
$\overline{126}^d$ are scalar fields getting SU(2) breaking vevs
that contribute to $M_D$ and $M_E$.

The entries $A$, $B$, $C$ of the up matrix can each arise from couplings to 10
or $\overline{126}$ scalar mesons.\footnote{The 120 scalar meson
would lead to antisymmetric contributions to the mass matrix, and
is therefore not needed.} There may be several such scalar mesons,
distinguished by discrete symmetries necessary to ensure the
texture structure of eq.(1). Each $\overline{126}$ contains an
SU(5) preserving vev that can contribute to the Majorana mass
matrix $M_{NN}$ of the SU(5) singlet right handed neutrinos.
To proceed further and make predictions for neutrino masses, we
need to introduce some hypothesis that limits the number of
parameters in the theory. We already know that we will need to
introduce at least one additional parameter into the theory,
namely the overall scale of the right handed neutrino masses.  To
maximize predictive power we will seek theories where this is the
{\em only} additional parameter.  This has some implications for
both the Yukawa couplings and the vevs:

(I) There are no new Yukawas, i.e., the Yukawa couplings giving
rise to $M_{NN}$ are the same as those that give rise to $A,\; B,\;
C,\; D,\; E,\; F$.

(II) All the entries in the $M_{NN}$ matrix must be generated
from the vev of only {\em one} of the $\overline{126}$ multiplets.

(III) Each fermion mass matrix element is generated by the vev of
only {\em one} of the 10 or $\overline{126}$ multiplets.

It is now easy to see that these minimality hypotheses lead us to
a theory in which $M_U$ originates in the following Yukawa
couplings

$$16 \left( \begin{array}{ccc}
0 & c\overline{126}\,^{uN} & 0\\[5pt]
c\overline{126}\,^{uN} & 0 & bX^u\\[5pt]
0 & bX^u & a\overline{126}\,^{uN}
\end{array} \right)16 \eqno (4)$$

\noindent
where:

(a) $\overline{126}\,^{uN}$ gets an electroweak breaking vev
giving rise to Dirac masses for up quarks and neutrinos.

(b) $\overline{126}\,^{uN}$ also gets an SU(5) preserving vev
contributing to $M_{NN}$.

(c) $X^u$ can be either $10^u$ or $\overline{126}\,^u$.  In
either case it only gets a vev in an electroweak breaking
direction and gives Dirac masses to up quarks and neutrinos.

It is easy to see that if we try to deviate away from eqs.\ 3 and
4 or from (a), (b), (c) we would either unnecessarily increase the
number of parameters, or we would end up with a light right-handed
neutrino that would become the Dirac partner to $\nu_e$, $\nu_\mu$
or $\nu_\tau$, with a mass which is much larger than present
laboratory limits. It is straightforward to find a set of
symmetries which guarantees the textures of equations (3) and (4).
Note that $X^u$ cannot be $\overline{126}\,^{uN}$; if it were,
wavefunction mixing would induce a non-zero 22 entry in the matrix.

The neutrino masses which follow from equation 4 are:

$$M_{\nu N} = \left( \begin{array}{ccc}
0 & -3C & 0\\
-3C & 0 & -3\kappa B\\
0 & -3\kappa B & -3A
\end{array} \right){v\over \sqrt{2}}\sin\beta$$\\
$$
M_{N N} = \left( \begin{array}{ccc}
0 & C & 0\\
C & 0 & 0\\
0 & 0 & A
\end{array} \right)V \eqno (5)$$
\\[-.2in]

\noindent
where $\kappa = 1$ if $X^u = \overline{126}\,^u$ (case I) and
$\kappa = - {1 \over 3}$ if
$X^u = 10^u$ (case II).  $V$, the superheavy singlet vev, is  the
one additional free parameter which occurs in the neutrino mass
matrix.

There are two important ingredients we left out of our discussion
so far:

1) All the quantities involved in the mass matrices are complex.
This appears to limit the predictive power; however all but one of
the phases can be eliminated by rephasing the fields.

2) The mass matrices that we measure at the weak scale are not
the same as those at the GUT scale (eqs.\ 1, 5); the two are
connected via the renormalization group [RG].

In our previous paper\cite{dhr}, we analytically solved the RG
equations for quark and charged lepton masses.  We shall restrict
ourselves in this paper to the same approximations used there.
We use one loop RG equations, neglecting all Yukawa couplings
except for the top, $\lambda_t$.  This is a good approximation only
for sufficiently small $\tan\beta$. The effects of large $\tan\beta$
will be studied in a forthcoming paper \cite{gregetal}.

The neutrino mass derives from effective dimension 5
operators, involving lepton doublets $L_i$ and Higgs doublet $H$,
of the form

$${{M^{ij}_{\nu\nu}} \over 2} L_i L_j \left({H\over v\sin\beta/\sqrt{2}}
\right)^2
$$
\noindent
where
$$M_{\nu\nu} = M_{\nu N}\, M^{-1}_{NN}\, M_{\nu N}^T$$

\noindent
$M_{\nu\nu}$ gets rescaled by an overall factor, as a result of
RG running from $M_G$ to $M_W$.
These ingredients result in the following mass matrices at the
weak scale:

$$
M_E = \eta_e \left( \begin{array}{ccc}
0 & Fe^{i\phi} & 0\\
Fe^{-i\phi} & -3E & 0 \\
 0 & 0 & D
\end{array} \right)  {v\over\sqrt{2}} \cos \beta
\eqno (6)
$$
$$
M_{\nu\nu} = \eta_\nu {9A v^2 \over 2V}\left( \begin{array}{ccc}
0 & C/A & 0\\
C/A & \kappa^2\, B^2/A^2 & 2\kappa\, B/A\\
0 & 2\kappa\, B/A & 1
\end{array} \right)\; \sin^2 \beta\,
$$
where $\eta_e$ and
$\eta_\nu$ take into account the RG scaling from $M_G$ to $M_W$.
Note that, apart from the overall scale, the neutrino mass matrix
depends only on $\kappa$, B/A, and C/A. As always A,B and C refer
to parameters renormalized at the GUT scale.

Harvey, Ramond and Reiss\cite{harvey} have discussed neutrino mass matrices
in SO(10) GUTs which incorporate the Georgi-Jarlskog ansatz. However, they
did not make any predictions for the neutrino masses. This is because,
even though our ansatz is a special limit of theirs, we rely on
the additional hypothesis that the vev of only one multiplet
contributes to any one matrix element (item III above). It is this
additional hypothesis which gives our ansatz its strong predictive
power with all neutrino masses and mixing angles determined, up to one overall
scale, in terms of parameters fixed in the charged fermion sector
of the theory.

The mass matrices in the lepton sector, Eq. 6, can be
diagonalized by bilinear transformations of the form:
\begin{eqnarray*}
M_E^{diag} & = & V^L_e\, M_E\, V_e^{R \dagger}\\[5pt]
M_{\nu\nu}^{diag} & = & V_{\nu}\, M_{\nu\nu}\, V_{\nu}^T
\end{eqnarray*}

\vspace{-.8in}
$$\eqno (7)$$
\\
\noindent
The leptonic CKM matrix is:
$$V' = V_{\nu}\; V_e^{L \dagger} \eqno (8)$$

\noindent
The natural parameterization for $V_e^L$, $V_{\nu}$ and $V'$ are:

$$V_e^L = \left( \begin{array}{rrr}
c_1^\prime & -s_1^\prime & 0\\[5pt]
s_1^\prime & c_1^\prime & 0\\[5pt]
0 & 0 & 1 \end{array} \right)
\left( \begin{array}{rrr}
1 & &\\[5pt]
&e^{i\phi} & \\[5pt]
& & e^{i\phi} \end{array} \right )$$
\\
$$V_{\nu} = \left( \begin{array}{rrr}
c_2^\prime & s_2^\prime &0\\[5pt]
-s_2^\prime & c_2^\prime & 0\\[5pt]
0 & 0 & 1 \end{array} \right)
\left( \begin{array}{rrr}
1 & 0 & 0\\[5pt]
0 & c_3^\prime & s_3^\prime\\[5pt]
0 & -s_3^\prime & c_3^\prime \end{array} \right) \eqno (9)$$
\\
$$V' = \left( \begin{array}{ccc}
c_1^\prime c_2^\prime - s_1^\prime s_2^\prime e^{-i\phi} &
s_1^\prime + c_1^\prime s_2^\prime e^{-i\phi} & s_2^\prime s_3^\prime\\[5pt]
-c_1^\prime s_2^\prime - s_1^\prime e^{-i\phi} &
-s_1^\prime s_2^\prime + c_1^\prime c_2^\prime c_3^\prime e^{-i\phi} &
s_3^\prime\\[5pt]
s_1^\prime s_3^\prime & -c_1^\prime s_3^\prime & c_3^\prime e^{i\phi}
\end{array} \right)$$

\noindent
The angles are given by:
\begin{eqnarray*}
s_1^\prime  &  =&  - {\frac{F}{3E}}\\[5pt]
s_2^\prime & = & {C/A \over {3\kappa^2 \; B^2/A^2}} > 0\\[5pt]
s_3^\prime & = & -2\kappa \; B/A
\end{eqnarray*}
\vspace{-.9in}
$$\eqno (10)$$
\\[.1in]
and the CP violating angle $\phi$ is determined in the previous
paper [1].

In general $V'$ contains three independent phases.
However, for our case all three are related to the phase $\phi$
which is identical to that occurring in the quark sector, and is
determined to be cos $\phi = .38 {{+ \ .21}\atop{- \ .14}}$.
Diagonalization of the quark mass matrices leads to a KM matrix
$V$ which is the same function of angles $\theta_i, \phi$ as $V'$
is of $\theta'_i$ and $\phi$:
$V'(\theta'_i, \phi) = V(\theta_i, \phi )$.
The relations between the mixing angles in the quark and lepton
sectors just involve simple group theory numerical factors
$$
\eqalignno{
s'_1 &= - {1\over 3} \ s_1\cr
s'_2 &= {1\over 3\kappa^2} \ s_2\cr
s'_3 &= 2\kappa \eta_3 \ s_3&(11)\cr}
$$
except for $\eta_3 = \eta_c V^2_{cb} m_t/m_c$ which comes from the effect of
the large top Yukawa coupling on the RG scaling of $V_{cb} = s_3$.
While the mixing angle from the D/E sector is most accurately
determined by
$$
s'_1 =- {\sqrt{ {m_e\over m_\mu}}}\eqno(12)
$$
the angles in the $U/\nu$ sector must be determined from quark
physics via:
$$
\eqalignno{
s_2 &= \sqrt{{m_u\over m_c}}\cr
s_3 &= V_{cb}.&(13)
\cr}
$$

We now use the known numerical inputs from the $U/D/E$ sectors to
make precise numerical  predictions of the neutrino masses.
The neutrino mass ratios are
$$
{m_{\nu_\tau}\over m_{\nu_\mu} } = {1 \over 3\kappa^2} \left( {B\over
A}\right)^{-2} = {1 \over 3\kappa^2} (\eta_3 V_{cb})^{-2} =
\cases{ 208 \pm 42  &(I)\cr
1870 \pm 370 &(II) \cr}\eqno(14)
$$
and
$$
{m_{\nu_\mu}\over m_{\nu_e}} = \left({C/A\over 3\kappa^2
B^2/A^2}\right)^{-2} = 9\kappa^4 \left( {m_u\over m_c}\right)^{-1} =
\cases{ 3100  \pm 1000  &(I)\cr 38 \pm  12  &(II)\cr}\eqno(15)
$$
where the ranges obtained correspond to ranges of the input parameters\\
$(V_{cb}, m_b, m_c, \alpha_s, m_u/m_d)$ which we have found to be consistent
with the quark mass and mixing predictions of our scheme [1].  We
cannot predict the overall mass scale.  We find
$$ m_{\nu_\tau} =  \eta_\nu {9A v^2 \over 2V}
\sin^2 \beta = .8 eV \left({m_t \over 170 GeV}\right)
\left({10^{14} GeV \over V}\right)
$$
where we have ignored electroweak contributions to $\eta_\nu$.

The most useful form for the prediction of the three mixing angles
is in terms of the predictions for neutrino oscillations from
flavor $i$ to flavor $j$. For sufficiently large $\Delta m_{ij}^2$
the oscillation probabilities can be approximated by the familiar
$2 \times 2$ case: $P_{ij} = \frac{1}{2} sin^2 2\theta_{ij}$.
We find
$$\theta_{\mu\tau}\simeq V'_{\nu_\mu\tau} = s'_3  ,$$
$$\theta_{e\mu}  \simeq |V'_{\nu_e\mu}| = \left(\frac{m_e}{m_\mu} +
\frac{m_{\nu_e}}{m_{\nu_\mu}} - 2 \sqrt{\frac{m_e m_{\nu_e}}{m_\mu
m_{\nu_\mu}}} \cos \phi \right)^{1/2} $$ and
$$
\theta_{e\tau} \simeq V'_{\nu_\tau e}= s'_1s'_3  .
$$
Using Equations 11-13, we obtain
$$
sin^2 2\theta_{\mu\tau} = \cases{ (2.6 \pm 0.5) 10^{-2} &(I)\cr
(2.9 \pm 0.6)10^{-3} &(II)\cr}\eqno(16)
$$
$$
sin^2 2\theta_{e\mu} = \cases{ (1.7 \pm 0.2) 10^{-2} &(I)\cr
(9.0 \pm 4.3) 10^{-2}  &(II)\cr}\eqno(17)
$$
$$
sin^2 2\theta_{e\tau} = \cases{(1.3 \pm 0.3) 10^{-6} &(I)\cr
(1.4 \pm 0.3)10^{-7} &(II).\cr}\eqno(18)
$$

The best hope for an experimental laboratory test of these numbers
is provided by $\nu_\mu \nu_\tau$ oscillations.
Present limits and the reaches of proposed experiments are shown
in the $\Delta m^2-sin^2 2\theta$ plot in Figure 1, together with
our two predictions as vertical lines.

In model I $\theta_{\mu\tau}$ is sufficiently large that the Fermilab E531
results imply that $m_{\nu_\tau} \leq 2.5$ eV. This
means that it is unlikely that planned neutrino oscillation
experiments\cite{p803,chorus,nomad} will be able to detect the neutrino
masses of this model. Given the upper bound on $m_{\nu_\tau}$ we
have, using Eq. 14, $m_{\nu_\mu} < 1.5 \times 10^{-2}$ eV or
$m^2_{\nu_\mu} < 2.3 \times 10^{-4}$ (eV)$^2$. Now given this upper
bound on $m_{\nu_\mu}$ and our value for $\theta_{e\mu}$, Eq. 17, (see
vertical line labelled I in figure 2) we find a possible resolution of the
Cl, Kamiokande and Gallium\cite{davis,kam,sage,gallex} solar neutrino
experiments by MSW oscillations, at the 90 \% confidence level.  Our
value of $\theta_{e\mu}$ implies that, as the error bars on the Ga
experiments\cite{sage,gallex} are decreased, a low number of about
50$\pm10$ SNUs will result. To test this region of parameter space in the
lab would require a long baseline $\nu_\mu \nu_\tau$ oscillation search
with sensitivity to smaller mixing angles than the present proposals.
We note that the neutrinos in this solution are all too light to be a
significant component of the dark matter.

In model II $\theta_{\mu\tau}$ is just beyond the E531 limits. This is very
exciting because it means that the upcoming $\nu_\mu \nu_\tau$ oscillation
searches will probe a large range of $\Delta m^2$ in this
model\cite{p803,chorus,nomad}. In particular, if $\nu_\tau$ makes a
significant contribution to the dark matter in the universe, then O(50)
events will be seen and $\sin^2 2\theta_{\mu\tau}$ will be determined to
be within 15\% of 3.10$^{-3}$.  We can still obtain an upper limit on
$m_{\nu_\tau}$ if we  demand that $\nu_\tau$ doesn't
overclose the universe.  We have $$m_{\nu_\tau} \leq 93 eV (\Omega_\nu
h^2) $$ with the Hubble constant $H_0 = 100 h km/s/Mpc$ and $ 1/2 \leq h
\leq 1$.  For $h = 1/2$ and $\Omega_\nu = 1$ , we have $m_{\nu_\tau} < 23
$ eV.  This implies, using Eq.14, $m_{\nu_\mu} < 1.5 \times 10^{-2}$ eV
or $m^2_{\nu_\mu} < 2.4 \times 10^{-4}$ (eV)$^2$.

Now consider possible MSW oscillations.  We find two possible
solutions to the solar neutrino problem consistent with the ``cosmological"
upper bound on $m_{\nu_\mu}$ and our value for $\theta_{e\mu}$, Eq. 17,
(depicted as the vertical line labelled II in figure 2). In fact, if a
signal is seen in $\nu_\mu \nu_\tau$ oscillations at CERN or at Fermilab
which is consistent with our value of $\theta_{\mu\tau}$, then we predict
values of $\theta_{e\mu}$  and $m_{\nu_\mu}$  which are significant for
the Cl, Kamiokande and Gallium solar neutrino experiments.   This is our
upper  model II solution. In this case we find about $110$ SNUs in Ga.
Note, this region is not consistent with combined fits to the
present observations of solar
neutrinos at 90 \% CL\cite{davis,kam,sage,gallex}.  Nevertheless, perhaps
it is too early to rule it out.\footnote{It might, in fact, be consistent
with the solar model with a 5\% higher core temperature\cite{bludman}.}
It is an interesting solution since, in addition to having a tau neutrino
with mass of order 20 eV and thus a significant component of the dark
matter\footnote{This could lead to a mixture of cold dark matter,  some
type of neutralino, and hot dark matter which seems to be preferred by
recent COBE data\cite{cobe}},  it also provides a possible solution to
the supernova shock reheating problem.  The values of $m_{\nu_\tau}$ and
$\theta_{e\tau}$ are just in the range required by Fuller et.
al.\cite{fuller} to allow for $\nu_\tau \nu_e$ oscillations in the
supernova.  This results in higher $\nu_e$ energies behind the shock,
thus producing an increase in the heating rate.

If no signal is seen in $\nu_\mu \nu_\tau$ oscillations at CERN or
at Fermilab, then the neutrino mass limits are sufficiently suppressed
that there is no hope that $\nu_\mu \nu_e$ oscillations could be found
at subsequent experiments such as the long baseline proposal at
Fermilab (P822)\cite{p822}. Even if a signal were seen in $\nu_\mu
\nu_\tau$ oscillations at CERN, a subsequent signal could only be seen at
experiments such as proposed in Fermilab P822 if the $\nu_\tau$ mass were
above the cosmological limit of $\sim$ 23 eV; a limit which we would expect to
apply to $\nu_\tau$ in this theory.

A third possible MSW solution to the solar neutrino problem is seen in
figure 2 as the lower model II solution.  It is not favored by GALLEX,
but is consistent with Chlorine and Kamiokande.

In conclusion, we have a very predictive model for neutrino masses and
mixing angles. In terms of just one arbitrary parameter, the overall scale
of the neutrino masses, we predict all 9 observable quantities of the
neutrino sector. In version I of our model, we find the value of
$\theta_{e \mu}$ and the allowed values for $m_\mu$ leads to $\nu_e
\nu_\mu$ resonant MSW neutrino oscillations which seems to be favored by
present experiments to solve the solar neutrino problem. In this case we
predict that with greater statistics the GALLEX and SAGE experiments will
settle on a result of around 50 SNUs. On the other hand, in version II of
our model a large region of parameter space will be probed by the NOMAD
and CHORUS experiments for $\nu_\mu \nu_\tau$ oscillations under
construction at CERN or by P803 proposed at Fermilab. For a sufficiently
large $\nu_\tau$ mass, including values for which the $\nu_\tau$
contributes a significant amount of dark matter to the universe, these
experiments will make a precision test of our theory, measuring sin$^2 2
\theta_{\mu\tau}$ to 15\% accuracy. This model is also relevant for solar
neutrino experiments, but is in a region of the $\Delta m^2 - \sin^2
2\theta_{e\mu}$ which is not consistent with the present data at 90\%
CL.  Our three possible MSW solutions to the solar neutrino problem are
given in table I, along with some of their properties.

S.R. thanks the Aspen Center For Physics where part of this work was carried
out.

\newpage

\vspace{.4in}
\begin{center}
\begin{tabular}{|c|c|c|c|}
\multicolumn{4}{c}{Table 1}\\ \hline
Model & I & II$_{upper}$ & II$_{lower}$  \\ \hline
$m_{\nu_\mu} $ & $1.7 \times 10^{-3}$ eV & $10^{-2}$ eV & $6 \times
10^{-4}$ eV   \\[5pt]
$m_{\nu_\tau} $ &  0.4 eV & 20 eV & 1 eV \\[8pt]
SAGE/Gallex &&&\\ \cline{1-1}
Counting Rate & $\sim 50$ SNU & $\simeq 110$ SNU & $\le 10$ SNU
\\[8pt]
$\nu_\mu - \nu_\tau$ osc. &&&\\ \cline{1-1}
$L_{\nu_\mu\nu_\tau}$ & 100 km & 44 m & 16 km \\[5pt]
$\sin^2 2\theta_{\mu\tau}$ & .026 & .003 & .003  \\ \hline
\end{tabular}
\end{center}

\vspace{.2in}
\noindent
\underline{Table Caption}:  The 3 possibilities for neutrino
masses in the case where the solar neutrino problem is solved via
MSW.  The $\nu_\mu-\nu_\tau$ oscillation lengths are for
$E_{\nu_\mu} = 20$ GeV.

\newpage

\vspace{.2in}
\begin{large}
\noindent
\underline{\bf Figure Captions}
\end{large}
\vspace{.2in}

\noindent
\underline{Figure 1}:  Present and future (P803, CHORUS, NOMAD and P860)
limits on the
$\nu_\mu-\nu_\tau$ mixing angle and $\Delta m^2_{\nu_\mu-\nu_\tau}$.
Taken from the addendum to the Fermilab P860 proposal, June 1992.
The vertical lines labeled I and II are our predictions for $\kappa=1$ and
$\kappa =-{1 \over 3}$ respectively.\\

\noindent
\underline{Figure 2}:  The region in $\nu_e - \nu_\mu$ mixing and $\Delta
m^2_{\nu_e-\nu_\mu}$ relevant for the MSW solution to the solar neutrino
problem. The 90\% C.L. regions are shown for the Homestake (dotted region),
Kamiokande II+III (solid line) and GALLEX (dashed line) experiments. The shaded
region is the combined fit of these three experiments (90\% C.L.). This figure
is reproduced from reference 9.
The vertical lines labeled I and II are our predictions for $\kappa=1$
and $\kappa =-{1 \over 3}$ respectively.\\

\noindent
\underline{Figure 3}: The $\nu_e$ survival fraction contours (solid lines)
for Gallium solar neutrino experiments as a function of $\nu_e - \nu_\mu$
mixing and $\Delta m^2_{\nu_e-\nu_\mu}$. The region allowed at 90\% C.L. by
the GALLEX experiment is shaded. This figure is reproduced from reference 9.
The vertical lines labeled I and II are our predictions for $\kappa=1$
and $\kappa =-{1 \over 3}$ respectively.\\
\end{document}